\documentclass[letter]{jpsj2}
\def\lsim{\buildrel {\textstyle <}\over {_\sim}}

\title{
$Z_2$-vortex ordering of the triangular-lattice Heisenberg antiferromagnet
}
\author{Hikaru Kawamura, Atsushi Yamamoto and Tsuyoshi Okubo}
\inst{Department of Earth and Space Science, Faculty of Science,
Osaka University, Toyonaka 560-0043,
Japan}
\recdate{\today}
\abst{
 Ordering of the classical Heisenberg antiferromagnet on the triangular lattice is studied by means of a mean-field calculation, a scaling argument and a Monte Carlo simulation, with special attention to its vortex degree of freedom. The model exhibits a thermodynamic transition driven by the $Z_2$-vortex binding-unbinding, at which various thermodynamic quantities exhibit an essential singularity. The low-temperature state is a ``spin-gel'' state with a long but finite spin correlation length where the ergodicity is broken topologically.  Implications to recent experiments on triangular-lattice Heisenberg antiferromagnets are discussed.
}
\kword{%
frustration, triangular lattice, $Z_2$ vortex, topological transition, Monte Carlo simulation}
\begin{document}
\maketitle

 Ordering of geometrically frustrated magnets has attracted much recent interest. Antiferromagnetic Heisenberg model on the two-dimensional (2D) triangular lattice is a typical example of such geometrically frustrated magnets. Interest in the ordering of the model has been enhanced by recent experiments on various triangular-lattice Heisenberg antiferromagnets (AFs) including, $S$=3/2 NaCrO$_2$ \cite{Olariu,Hsieh}, $S$=1 NiGa$_2$S$_4$ \cite{Nakatsuji,Takeya,Yaouanc,MacLaughlin,Yamaguchi} and $S$=1/2 organic compounds, $\kappa$-(BEDT-TTF)$_2$Cu$_2$(CN)$_3$ \cite{Shimizu} and EtMe$_3$Sb[Pd(dmit)$_2$]$_2$ \cite{Itou}. While most of these compounds exhibit a spin-liquid-like behavior at low temperatures without the conventional magnetic long-range order (LRO), all of them turn out to exhibit a weak but clear transition-like anomaly at a finite temperature. This ``transition'' seems intrinsically 2D, being neither the standard AF nor the spin-glass transition. Thus, it remains most interesting to clarify the nature of the ordering process of the frustrated AF Heisenberg model on the triangular lattice.

 It is now established that the Heisenberg AF on the triangular lattice interacting via the nearest-neighbor bilinear interaction exhibits a magnetic LRO at $T=0$, the so-called 120-degrees structure, in either case of quantum $S=1/2$ spin \cite{Bernu,Capriotti} or classical $S=\infty$ spin. Because of the two-dimensionality of the lattice, the magnetic LRO is established only at $T=0$, while the associated  spin correlation length diverges exponentially toward $T=0$. Some time ago, it was demonstrated by one of the present authors (H.K.) and Miyashita that the model bears a topologically stable point defect characterized by a two-valued topological quantum number, a $Z_2$ vortex \cite{KawaMiya}. 
Kawamura and Miyashita suggested that, in contrast to its unfrustrated counterpart, the triangular Heisenberg AF might exhibit a thermodynamic phase transition at a finite temperature driven by the binding-unbinding of the $Z_2$ vortices, where the vortex correlation length $\xi_v$, corresponding to the mean separation of free $Z_2$ vortices, diverges {\it keeping the spin correlation length $\xi$ finite\/}.

 It should be stressed that the proposed vortex transition is of different character from the Kosterlitz-Thouless (KT) transition of the 2D {\it XY\/} model in that there appear two distinct length scales, or two different stiffnesses (energy scales), in the former: One is a vortex correlation length $\xi_v$ diverging at the vortex transition $T=T_v$, and the other is a spinwave correlation length $\xi_{sw}$ staying finite at $T=T_v$. 
Counterview to such a picture is that there is no such ``vortex-spinwave decoupling'' that the $Z_2$-vortex transition is merely a rapid crossover rather than a true thermodynamic transition \cite{APYoung}.
 
 Under such circumstances, we wish to investigate in the present Letter the nature of the $Z_2$-vortex ordering of the classical AF Heisenberg model on the triangular lattice, first analytically by use of mean-field and scaling analyses, and then numerically by use of a Monte Carlo (MC) simulation. On the basis of these analytical and numerical results, we shall discuss the recent experimental results on various triangular AFs.

 The model we consider is the classical Heisenberg AF on the two-dimensional triangular lattice, whose Hamiltonian is given by ${\cal H} = J \sum_{<ij>} \vec S_i\cdot \vec S_j$, where $J>0$ and the sum is taken over all nearest-neighbor pairs.

 As demonstrated in ref.\cite{KawaMiya}, the Heisenberg spins, or more precisely the chirality vectors, circulate around a vortex core making a topologically stable vortex, whereas whether they circulate in clockwise or counter-clockwise fashion does not make clear distinction topologically. At $T=0$, an isolated vortex of its radius $R$ has an energy $E_v \sim c\log R + \mu$, with $c$ an energy constant and $\mu$ the vortex core energy.  Stability of a single vortex against the entropic effect yields a rough estimate of the vortex transition temperature $T_v=c/2$ above which an isolated $Z_2$ vortex is spontaneously generated \cite{KawaMiya}. 

  Suppose that there exist $N_v$ free vortices on the lattice of size $N=L\times L$. Mean separation between free vortices is given by $\xi_v = 1/\sqrt{n_v}$ where $n_v\equiv N_v/N$ is the number density of free vortices. Neglecting the correlation effects between vortices, the free energy of an assembly of vortices might be given by $F_v\sim (c\ln \xi_v + \mu)N_v - T\ln [N!/(N_v!(N-N_v)!)]$. By minimizing $F_v$ with respect to $n_v$, we get $n_v\sim \exp [-(\mu-T_v)/(T-T_v)]$ with $T_v=c/2$ as above. Then, the vortex correlation length $\xi_v$ corresponding to the mean separation of free vortices is obtained as $\xi_v\approx \exp [\frac{\mu-T_v}{2(T-T_v)}]$.  Toward $T=T_v$, $\xi_v$ diverges exponentially, and free vortices disappear at $T<T_v$. Previous MC suggested $T_v\simeq 0.28$ (in units of $J$) \cite{KawaYama} and $\mu \simeq 1.65$ \cite{KawaMiya}.

 In the above mean-field analysis, we have neglected the correlation effect (or the screening effect) between vortices. In reality, the screening effect would make neighboring vortices getting closer in distance. In view of this, we generalize the mean-field expression of $\xi_v$ to the form
\begin{equation}
\xi_v\approx \exp[\left( A/(T-T_v) \right) ^\alpha],\ \ \ (T>T_v),
\end{equation}
where $A$ is a constant and the exponent $\alpha$ is expected to be smaller than the mean-field value unity, {\it i.e.\/}, $\alpha < 1$. While one has $\alpha=1/2$ in the standard KT transition, precise criticality of the $Z_2$-vortex transition has not been elucidated yet. In any case, since the standard 2D Coulomb-gas description is not directly applicable to the $Z_2$-vortex ordering, there seems to be no strong reason to believe that the exponent $\alpha$ is exactly $\alpha=1/2$. 

 From a general scaling argument, the singular part of the free energy around $T_v$ is given by $f_v\approx \xi_v^{-2}$. At $T=T_v$, $f_v$ exhibits an essential singularity. Though weak, an essential singularity is a well-defined singularity so that the expected $Z_2$-vortex transition is a thermodynamic transition, not just a rapid crossover nor a smeared transition. The specific heat is given by%
\begin{eqnarray}
C \approx \left(A/(T-T_v) \right) ^{2(\alpha+1)} \exp[-2\left( A/(T-T_v) \right)^\alpha] 
\nonumber \\
 +\ [{\rm regular\ part}],\ \ \ (T>T_v),
\end{eqnarray}
where the regular part represents the contribution from degrees of freedom other than free vortices such as spinwaves. While the specific heat exhibits only a weak essential singularity at $T=T_v$ where the first vortex-pair unbinds, it would exhibit a non-singular peak or a shoulder at a temperature slightly above $T_v$ where the vortex-pair unbinding occurs most extensively. The precise form of the specific heat, however, would largely depend on the form of the regular part, which is a non-universal property sensitive to the details of each particular system.

 Spin correlation function $C(r_{ij})=<\vec S_i\cdot \vec S_j>$ is affected both by vortices and spinwaves. We assume here that the spin correlation can be factorized into the vortex part $C_v(r)$ and the spinwave part $C_{sw}(r)$, $C(r)\approx C_v(r)C_{sw
}(r)$, as in the KT ordering of the 2D {\it XY\/} model \cite{Jose}. Assuming the standard exponential decay for each correlation function at long distances, $C(r)\approx \exp(-\frac{r}{\xi})$, $C_v(r)\approx \exp(-\frac{r}{\xi_v})$ and $C_{sw}(r)\approx \exp(-\frac{r}{\xi_{sw}})$, we get $\xi=\xi_v\xi_{sw}/(\xi_v+\xi_{sw})$. Note that the vortex correlation length $\xi_v$ diverges at $T=T_v$, whereas the spinwave correlation length $\xi_{sw}$ stays finite at $T=T_v$, which monotonically increases with decreasing $T$ and eventually diverges toward $T=0$. In the high-temperature regime $T>>T_v$ where $\xi_v<<\xi_{sw}$, spin correlation is dominated by vortices, while in the low-temperature regime $T<<T_v$, dominantly by spinwaves. 
At temperatures slightly above $T_v$ where $\xi_v>>\xi_{sw}$, $\xi$ is given by
\begin{equation}
\xi =\xi_v\xi_{sw}/(\xi_v+\xi_{sw}) \sim  \xi_{sw}(1-\xi_{sw}/\xi_v)\ \ \ (T>T_v),
\end{equation}
with $\xi_v$ given by Eq.(1). Hence, $\xi$ exhibits a weak essential singularity at $T=T_v$, remaining finite there. The relative importance of vortex and spinwave excitations is interchanged around a crossover temperature $T_{\times} > T_v$ at which $\xi_v=\xi_{sw}$. The appearance of two distinct length scales $\xi_v$ and $\xi_{sw}$,  or two distinct coupling constants (stiffnesses), is a unique feature of the $Z_2$-vortex ordering of the Heisenberg system. This is in contrast to the standard KT ordering of the {\it XY\/} system where there exist only one length scale (one stiffness). 

 The low-temperature phase realized below $T_v$ is an ``ordered'' state in the sense that it cannot be reached adiabatically from the high-temperature phase, though it is a ``spin paramagnetic'' state with a finite spin correlation length. It is essentially a spinwave state where the full ergodicity is broken {\it topologically\/} since only the vortex-free sector is allowed on long length scale in the phase space. Dominance of spinwave excitations below $T_v$ would give rise to the specific heat proportional to $T^2$ at $T<T_v$, once the quantum effect has been taken into account (in the pure classical model, the $T^2$ specific heat is not realized) \cite{Fujimoto}. Although the spin correlation length $\xi$ or the spin correlation time $\tau$ still remains finite even below $T_v$, the suppression of free vortices might lead to a substantial growth of $\xi$ or $\tau$ at and below $T_v$, leading to a near-critical (but not true critical) state. We call such a topologically ordered state with a finite but long spin correlation length and correlation time a ``spin gel'' state. This state differs in nature either from the standard AF ordered state, from the spin-liquid state, or from the the spin-glass state.

 Finiteness of the spin correlation length $\xi$ at $T=T_v$ has important consequences on the response of the system against weak perturbative interactions. For example, a magnetic field, which couples to the spin via the Zeeman term, reduces the Hamiltonian symmetry so that one may suspect that even an infinitesimal magnetic field suppresses the $Z_2$-vortex transition. This is not necessarily the case, however, in the situation where $\xi$ stays finite at $T=T_v$. If $H$ is sufficiently small satisfying $H\xi(T=T_v)^2 \lsim k_BT_v$, the $Z_2$-vortex transition and the low-temperature ``spin gel'' state might remain essentially the same as in the zero-field ones. 

 Similar situation is expected for other perturbative interactions such as the magnetic anisotropy $D$ or the interplanar 3D coupling $J'$. If these perturbative interactions are weak enough satisfying $D\xi^2\lsim k_BT_v$ or $J'\xi^2\lsim k_BT_v$, the $Z_2$-vortex transition and the low-temperature spin-gel state might well persist. Thus, if $\xi$ stays sufficiently short at $T=T_v$, the $Z_2$-vortex transition and the spin-gel state of the 2D Heisenberg system might be ``protected'' against the weak anisotropy and/or the interplanar interaction.

 Under an infinitesimal field $H$, the singular part of the free energy still keeps the same form as the zero-field one where the transition temperature $T_v$ and the constant $A$ now weakly depend on $H$ but in a regular way, $T_v(H)=T_v(0)+cH^2+\cdots $ and $A(H)=A(0)+c'H^2+\cdots $. From this, prediction for the zero-field susceptibility $\chi$ follows,
\begin{eqnarray}
\chi \approx \left(A/(T-T_v) \right) ^{\alpha+1} \exp \left[ -2\left(A/(T-T_v)\right)^\alpha \right]
\nonumber \\
 +\ [{\rm regular\ part}],\ \ \ (T>T_v),
\end{eqnarray}
where the exponent $\alpha$ and the constant $A$ are common with the ones in Eqs.(1) and (2). Again, $\chi$ exhibits a weak essential singularity at $T=T_v$. Similar expression can also be derived for the nonlinear susceptibility $\chi_3$, where the exponent in the prefactor of Eq.(4) $1+\alpha$ should be replaced by $2(\alpha+1)$. 


 In order to examine the validity of the above analysis, we also perform a MC simulation  of the model based on the standard heat-bath method combined with the over-relaxation technique. The lattice is of $48 \leq L\leq1536$ with periodic boundary conditions. 

\begin{figure}[ht]
\begin{center}
\includegraphics[scale=0.5]{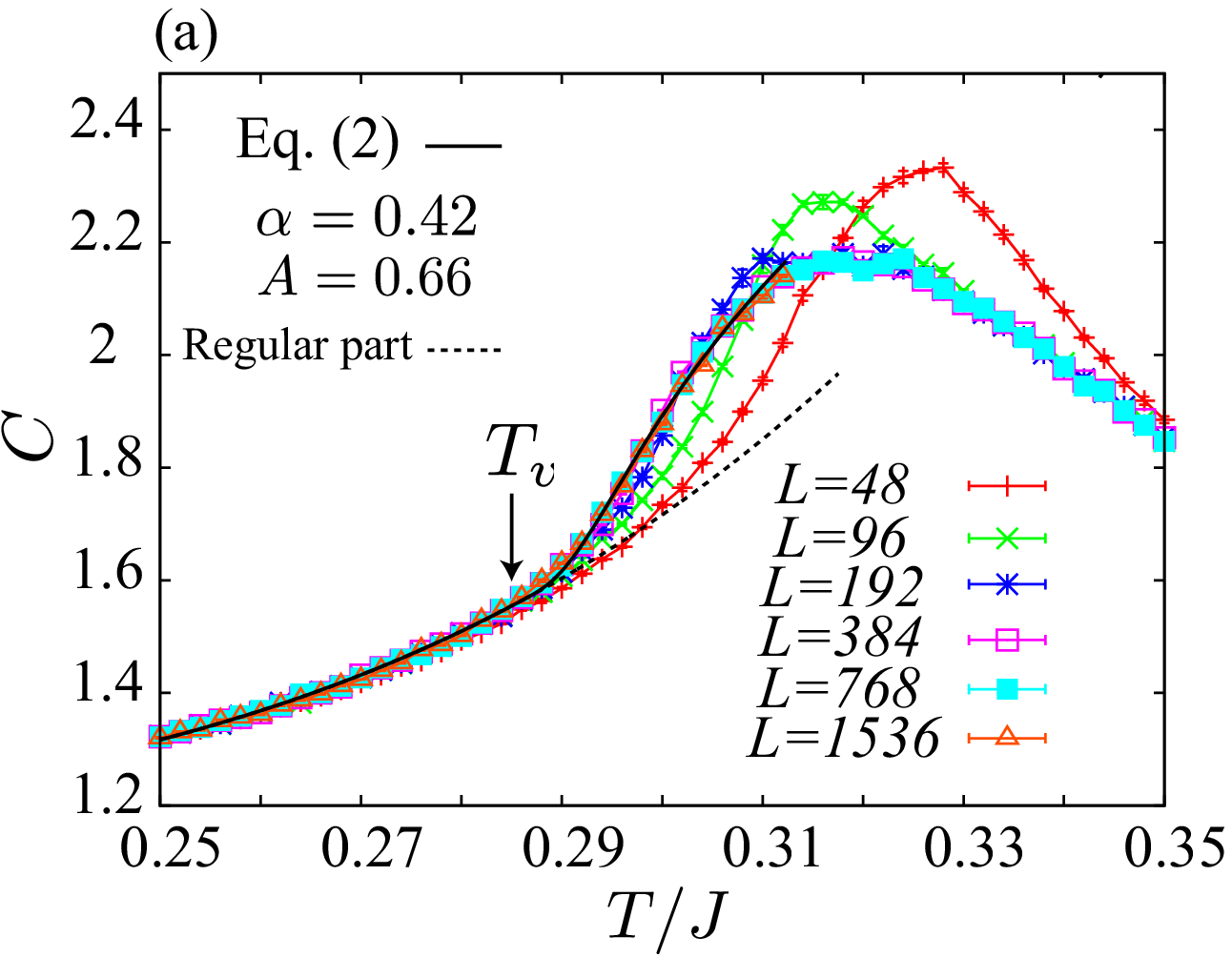}
\includegraphics[scale=0.5]{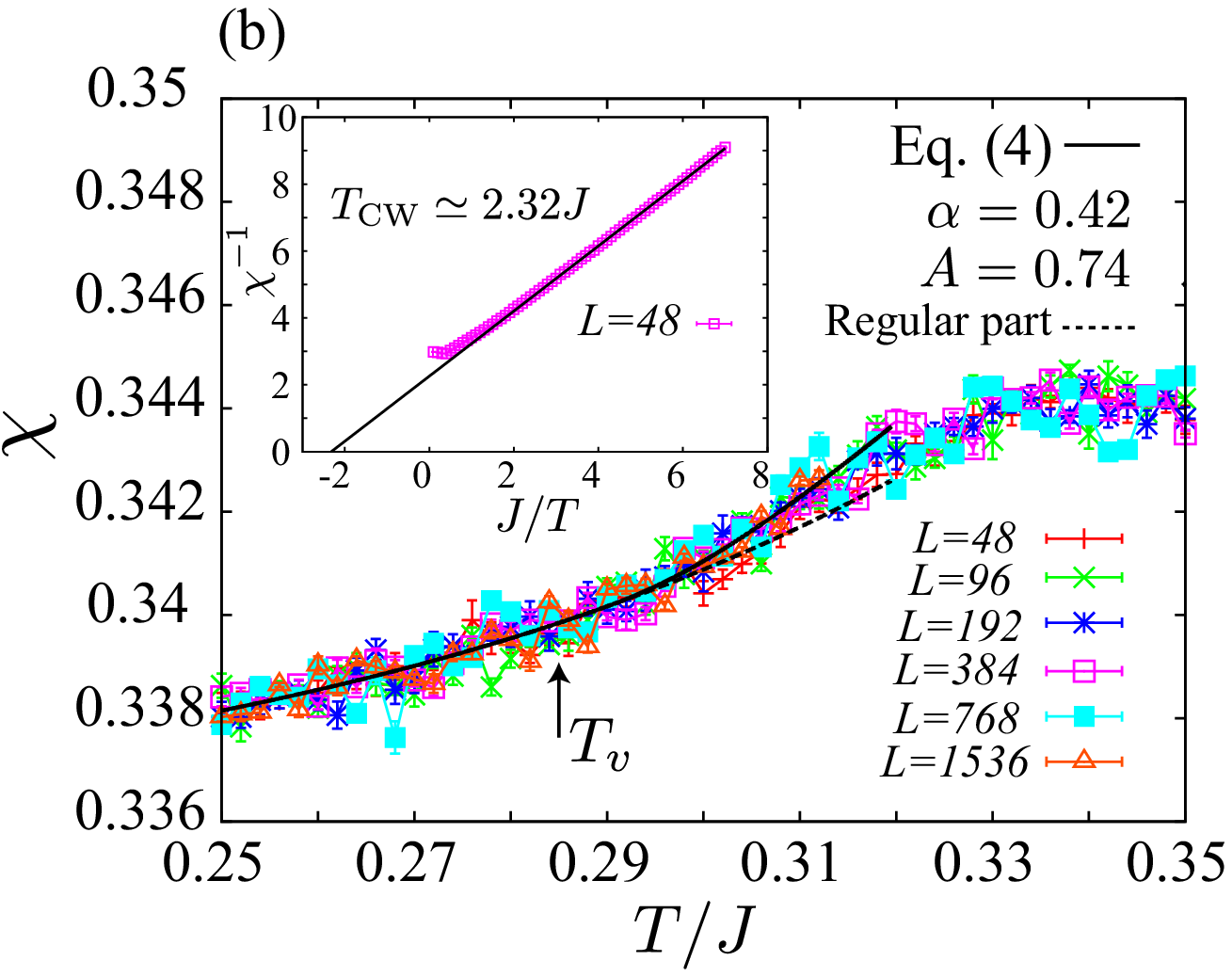}
\end{center}
\caption{
(Color online) Temperature dependence of the specific heat (a) and the susceptibility (b) per spin. The curves are the fits based on Eqs.(2) and (4). The inset is the Curie-Weiss plot for the inverse susceptibility.
}
\end{figure}

 The specific heat $C$ and the susceptibility $\chi$ are shown in Fig.1. The specific heat exhibits a distinct but rounded peak around $T\sim 0.318$, located slightly above $T_v$. More precise $T_v$-value will be determined below in Fig.3 as $T_v\simeq 0.285\pm 0.005$, together with the $\alpha$-value, $\alpha= 0.42\pm 0.15$. The Curie-Weiss temperature $T_{CW}$ is estimated as $T_{CW}\simeq 2.32$: See the inset of Fig.1(b). The $C$ data are then 
fitted to our theoretical expression Eq.(2) where the regular part is approximated by the fourth-order polynomial of $T$. As can be seen from Fig.1(a), the fit works well with $A=0.66\pm 0.05$. The $\chi$ data can also be fittable to Eq.(4) with $A=0.74\pm 0.55$, which agrees with the $A$ value  determined above from $C$.

 In Fig.2, we show on a semi-log plot the finite-size spin correlation length $\xi_L$ associated with the 120-degrees structure \cite{VietKawamura}. A scaling relation $\xi_{2L}/\xi_L\approx f(\xi_L/L)$ is expected to hold for larger $L$ \cite{Caracciolo}. As shown in the inset, the data turn out to scale well in this form for larger $L$, yielding a scaling function $f(x)$. Then, following ref.\cite{Caracciolo}, we extrapolate $\xi_L$ to the bulk correlation length $\xi=\xi_\infty$ on the basis of the above scaling relation, the resulting $\xi$ being given in Fig.2. The change in the behavior of $\xi(T)$ is discernible around $T=T_v$. The data can be well fitted to our theoretical expression Eq.(3) with $A\simeq 0.94\pm 0.4$ and $\alpha=0.42$ (fixed), where we fit $\ln \xi_{SW}$ by the fourth-order polynomial of $T$. Note that, in contrast to the $\xi$ data reported in ref.\cite{Wintel}, our $\xi$ data, or its temperature derivative, does not show any appreciable discontinuity at $T=T_v$, exhibiting only a weak singularity consistent with an essential singularity. This difference from ref.\cite{Wintel} probably comes from the fact that the two-loop RG formula used in ref.\cite{Wintel} in performing an $L=\infty$ extrapolation is no longer valid around $T=T_v$.
\begin{figure}[ht]
\begin{center}
\includegraphics[scale=0.6]{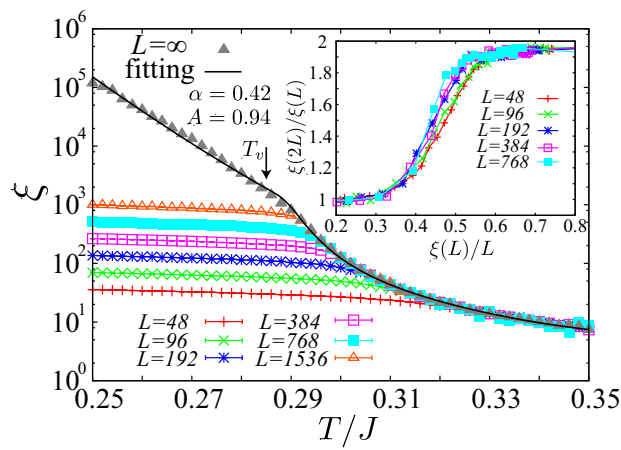}
\end{center}
\caption{
(Color online) Temperature dependence of the finite-size spin correlation length, together with the bulk one. The inset is a scaling plot, {\it i.e.\/}, $\xi_{2L}$/$\xi_L$ plotted vs. $\xi_L/L$.
}
\end{figure}

 In the present model, although the spin correlation length remains finite at $T=T_v$, it becomes quite large, $\xi\simeq 1700$. Our fitting analysis also gives an estimate of the crossover temperature $T_\times\simeq 0.294$ and the crossover length $l_\times \simeq 380$. It should be noticed, however, that the value of $\xi$ at $T=T_v$ (at $T=T_\times$) is a non-universal property depending on the details of each system. 

  In the previous studies, as an order parameter characterizing the $Z_2$-vortex transition, either a Wilson-loop (vorticity function)  \cite{KawaMiya} or a vorticity modulus \cite{KawaKiku} has been proposed. Here, we calculate  following Refs.\cite{Southern,KawaYama} the latter quantity, the vorticity modulus $v$, defined as the free-energy cost against a vortex formation $\Delta F$ divided by $\ln L$, $v=\Delta F/\ln L$. In Fig.3, we show the temperature derivative of the vorticity modulus, -${\rm d}v/{\rm d}T$, calculated from appropriate fluctuations. The data exhibit a sharp peak which sharpens with $L$, suggesting the occurrence of a phase transition.  In the inset, the peak temperature $T_{peak}(L)$ is plotted versus $1/\ln(L)$. Indeed, Eq.(1) implies a relation $T_{peak}(L)-T_v\approx (1/\ln L)^{1/\alpha}$. This fit yields $T_v=0.285\pm 0.005$ and $\alpha=0.42\pm 0.15$, which are the $T_v$ and $\alpha$ values quoted above. 

%
\begin{figure}[ht]
\begin{center}
\includegraphics[scale=0.6]{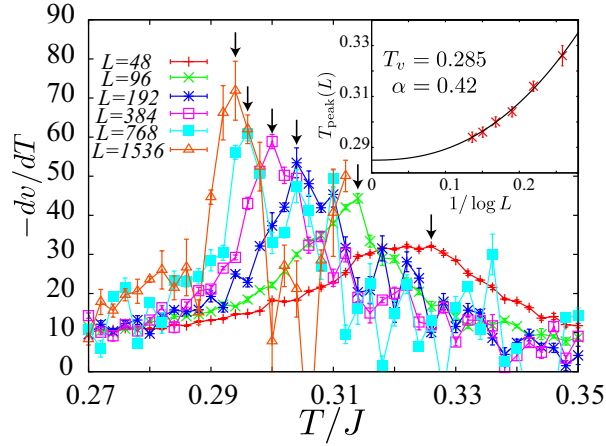}
\end{center}
\caption{
(Color online) Temperature dependence of the minus of the temperature derivative of the vorticity modulus. The inset represents the size dependence of the peak temperature.
}
\end{figure}
%


 We now wish to discuss possible implications of our results to recent experiments on several triangular-lattice Heisenberg antiferromagnets. We note $S$=3/2 NaCrO$_2$ ($T_{CW}\simeq$ 290K) \cite{Olariu,Hsieh} and  $S$=1 NiGa$_2$S$_4$  ($T_{CW}\simeq$ 80K) \cite{Nakatsuji,Takeya,Yaouanc,MacLaughlin,Yamaguchi} exhibit a strikingly similar ordering behavior in spite of the difference in their integer/half-integer spin quantum numbers. Indeed, NaCrO$_2$ (NiGa$_2$S$_4$) exhibit a clear but rounded specific-heat peak at $T_{peak}=41$K (10K), whereas a transition-like anomaly is observed at $T_f\simeq 30$K (8.5K), a temperature slightly below $T_{peak}$, where the spin dynamics is rapidly slowed down giving rise to a quasistatic internal field. However, the spin dynamics probed by  NMR, NQR, ESR and $\mu$SR is not completely frozen, but the spins remain slowly fluctuating even below $T_f$, unlike the conventional AF or the spin glass. Such a dynamically fluctuating ordered state extends over a wide temperature range, down to 10K (2K). The spin correlation length determined from neutron scattering is kept finite even at and below $T_f$, {\it i.e.\/}, $\xi\simeq 20$ (5) lattice spacings. At least in NiGa$_2$S$_4$, dynamical freezing at $T=T_f$ accompanies a weak structure in the susceptibility, while application of magnetic fields greater than $100\sim 1000$ G gradually changes the slowly fluctuating state into more conventional frozen state \cite{MacLaughlin}.

 These experimental features are fully consistent, at least qualitatively, with the $Z_2$-vortex order, if the experimental freezing temperature $T_f$ and the low-temperature state are identified as $T_v$ and the ``spin-gel'' state. In particular, energetics seems appropriate: Our simulation yields $T_{peak}/T_{CW}\simeq 0.137$ and  $T_v/T_{CW}\simeq 0.123$, which are close to the corresponding experimental values 0.14 and 0.10 (NaCrO$_2$), and 0.13 and 0.11 (NiGa$_2$S$_4$). Furthermore, in case of NiGa$_2$S$_4$, a weak anomaly observed in the specific heat and the linear and nonlinear susceptibilities are well fittable to our theoretical formula (2) and (4) \cite{Nambu}, while the crossover-field, $\sim 1000$G, estimated from our theoretical formula with the experimental value of $\xi\simeq 8$ seems consistent with the $\mu$SR measurement \cite{MacLaughlin}. Major quantitative discrepancy between experiments and our present simulation lies in the magnitude of the spin correlation length $\xi$ at the transition. In our simulation, $\xi$ is of order 1000 near $T=T_v$, while it is only 5 or 20 experimentally. So, some mechanism, not taken into account in the simplest classical Heisenberg model, is required to explain the shortness of $\xi$. As emphasized, the magnitude of $\xi$ is a nonuniversal property governed by the non-vortex physics, which could significantly be reduced, say, by quantum fluctuations, charge fluctuations, or further frustration effects associated with the interaction other than the main exchange coupling.

 $S$=1/2 organic AFs $\kappa$-(BEDT-TTF)$_2$Cu$_2$(CN)$_3$ and EtMe$_3$Sb[Pd(dmit)$_2$]$_2$ also exhibit a spin-liquid behavior without the conventional magnetic LRO down to low temperature. Recent measurements revealed a weak anomaly at a finite temperature, 4$\sim 6$K for BEDT-TTF \cite{Shimizu} and  $\sim$ 5K for dmit \cite{Itou}, where the NMR $T_1$ exhibits an anomaly (or the spectra broaden) and the specific heat exhibits a broad hump. We note that the $Z_2$-vortex order is one possible candidate of the experimentally observed anomaly. Of course, for the vortex to be a meaningful excitation, minimal amount of the noncollinear spin short-range order of a few lattice spacings is required, which, however, is not unlikely in these compounds. In order to substantiate this conjecture, further study is required.

The authors are thankful to S. Nakatsuji,  M. Hagiwara, Y. Nambu and H. Yamaguchi for discussion. This study was supported by Grant-in-Aid for Scientific Research on Priority Areas ``Novel States of Matter Induced by Frustration'' (19052006). We thank ISSP, Tokyo University for providing us with the CPU time.

\end{document}